# Interplay of quantum and real-space geometry in the anomalous Landau levels of singular flat bands

Xuanyu Long[1] and Feng Liu[1*]

[1]*Department of Materials Science and Engineering, University of Utah, Salt Lake City, Utah*

*84112, USA*

*To whom correspondence should be addressed: fliu@ eng.utah.edu

**Abstract**
Quantum geometry of electronic state in momentum space, distinct from real-space structural geometry, has attracted increasing interest to shed light on understanding quantum phenomena. An interesting recent study [Nature 584, 59-63 (2020)] has numerically solved a 2-band effective Hamiltonian to show the anomalous Landau level (ALL) spreading $\Delta$ of a singular flat band (SFB), such as hosted in a kagome lattice, in relation to the maximal quantum distance $d$ of the SFB, $\Delta(d)$, which enables a direct measure of quantum geometry. Here, we investigate the ALLs of SFB by studying both the 2-band Hamiltonian and a diatomic kagome lattice hosting two SFBs mirrored by particle-hole symmetry. We derive an exact analytical solution of the 2-band Hamiltonian to show there are two branches of $\Delta(d)$. Strikingly, for the diatomic kagome lattice, $\Delta$ depends on not only $d$ but also $r$, the real-space diatomic distance. As $r$ increases, $\Delta$ shrinks toward zero while $d$ remains intact, which can be intuitively understood from the magnetic-field-induced disruption of destructive interference of the SFB compact localized states. Based on semiclassical theory, we derive rigorously the dependence of $\Delta$ on $r$ that originates from the tuning of the non-Abelian orbital moment of the two SFBs by real-space geometry.

**Introduction**
Materials' physical behavior is critically governed by their structure-property relationships in which real-space geometry, especially lattice symmetry plays an important role. Remarkably, with the emergence of quantum materials in recent years, quantum geometry of electronic states in momentum space, a concept introduced four decades ago[1], has attracted much renewed interest, as it reveals new insights of quantum materials' exotic behavior governed by the quantum geometry-property relationships[2-9], distinctively beyond the conventional structure-property relationships. For example, it was shown before that the singular flat band (SFB) of a kagome lattice exhibits an anomalous Landau level (ALL) spectrum[10], beyond the Onsager's semiclassical quantization rule[11-15] that is widely used to explain the LL quantization under magnetic field. However, it is until recent that the ALLs of SFB is generally linked to quantum geometry[9]. A 2-band effective model Hamiltonian has been developed[9,16] to calculate and analyze the LL spectrum of SFB. Significantly, the numerical solution of the 2-band Hamiltonian reveals an empirical but quantitative dependence of the ALL spreading $\Delta$ on the maximal quantum distance $d$ of the SFB wavefunction, $\Delta(d)$, which enables a direct measure of quantum geometry[9].

Here, we investigate the LLs of SFBs by studying both the 2-band effective Hamiltonian[9] and the diatomic kagome lattice model that hosts two SFBs related by particle-hole symmetry, the so-called yin-yang flat bands[17,18] as shown in Fig. 1a, which has been experimentally realized in superatomic graphene and some organic frameworks[18,19]. We are able to derive an exact analytical solution of the 2-band Hamiltonian, and show there are actually two branches of $\Delta(d)$, including the one branch found by the numerical solution previously[9], which can be readily verified by reversing the magnetic field direction. Strikingly, for the diatomic kagome lattice, we found that $\Delta$ depends not only on $d$ but also on $r$, the real-space diatomic distance in the lattice, which shows *an intriguing interplay between quantum and real-space geometry in governing a quantum material's exotic behavior*. As $r$ increases, $\Delta$ gradually decreases until vanishing to zero at the maximal $r$, while $d$ remains the same. Intuitively, the newly revealed real-space geometry dependence of $\Delta$ is understood from the magnetic-field-induced disruption of destructive interference of the localized SFB wavefunction, namely the compact localized states (CLSs)[20]. Especially, at the maximal $r$, the destructive interference is reconstructed to be perfect under magnetic field, leading to a complete collapse of the ALLs. Furthermore, based on semiclassical theory, we derive rigorously that the dependence of $\Delta$ on $r$ originates from the non-Abelian orbital magnetic moment generated by the bipartite nature of diatomic kagome lattice with particle-hole symmetry, in analogy with the origin of electron spin from the Dirac equation[21-23].

The quantum geometry of one single band is defined via the infinitesimal Hilbert-Schmidt quantum distance[24-26] $dS$ between two Bloch states differed by an infinitesimal $d\boldsymbol{k}$: $dS^2 = 1 - |\langle\psi(\boldsymbol{k})|\psi(\boldsymbol{k}+d\boldsymbol{k})\rangle|^2$. The expansion coefficients of $dS^2$ with respect to $dk^\mu dk^\nu$, $Q_{\mu\nu}(\boldsymbol{k})$ is the quantum geometry tensor (QGT), where $\mu$, $\nu = x, y$ are the indices of momentum. The symmetric real part of QGT, $g_{\mu\nu}(\boldsymbol{k})$, is the Fubini-Study quantum metric[1,27], and the imaginary part is related to the antisymmetric Berry curvature tensor, $F_{\mu\nu} = -2\,\mathrm{Im}\,Q_{\mu\nu}$[28]. For SFBs, $g_{\mu\nu}$ diverges at the band crossing point (BCP) (see Fig. 1c). If one draws an infinitesimal closed loop around the BCP, the maximal possible quantum distance $d_{\max}$ between any two points on the loop characterizes the singularity of the FB wavefunction[9] (herein we emit the subscript of $d_{\max}$ for simplicity). For SFBs in Mielke's checkerboard and kagome lattices[29,30], $d = 1$ corresponding to the case of maximal singularity[9].

**Exact solution of 2-band Hamiltonian**
The low-energy physics of a SFB, such as hosted in a kagome lattice, can be generally captured by four independent parameters: three effective masses $M_{xx}$, $M_{yy}$ and $M_{xy}$ of the parabolic band (PB), and the maximal quantum distance $d \in [0,1]$ characterizing the strength of quantum singularity of the BCP[9,16]. Consider the isotropic case $M_{xx} = M_{yy}$ and $M_{xy} = 0$ (the more general anisotropic case can be obtained

by a simple coordinate transformation[9]), the 2-band continuum effective Hamiltonian reads

$$H_{\text{SFB}} = t \begin{pmatrix} k_x^2 + (1-d^2)k_y^2 & d\sqrt{1-d^2}k_y^2 - i\xi dk_x k_y \\ d\sqrt{1-d^2}k_y^2 + i\xi dk_x k_y & d^2 k_y^2 \end{pmatrix}. \quad (1)$$

$\xi = \pm 1$ denotes the chirality of the flat band (FB) wavefunction (see discussions below) and $t$ sets the energy scale. $k_x$ and $k_y$ are components of the lattice momenta $\boldsymbol{k}$. The solution of Eq. (1) gives two energy bands $E_0 = 0$ and $E_1 = t(k_x^2 + k_y^2)$ and their respective wavefunctions:

$$\boldsymbol{V}_0 = \frac{1}{\sqrt{k_x^2 + k_y^2}} \begin{pmatrix} -dk_y \\ i\xi k_x + \sqrt{1-d^2}k_y \end{pmatrix} \text{ and } \boldsymbol{V}_1 = \frac{1}{\sqrt{k_x^2 + k_y^2}} \begin{pmatrix} \sqrt{1-d^2}k_y - i\xi k_x \\ dk_y \end{pmatrix}. \quad (2)$$

When $0 < d < 1$, the FB wavefunction $\boldsymbol{V}_0$ has a chirality $\xi = \pm 1$ due to the sign of the imaginary part, just like the circular polarization of light. In the following we will solve $H_{\text{SFB}}$ analytically to obtain the exact expressions of ALL spectra.

Under magnetic field, we substitute $k_x$ and $k_y$ by the ladder operators $a$ and $a^\dagger$ satisfying $[a, a^\dagger] = 1$ in the standard way and define $l_B = \sqrt{\hbar/eB}$ as the magnetic length[9,31,32] [see Supplementary Information (SI)]. We then calculate the matrix elements of the substituted Hamiltonian $\widetilde{H}_{\text{SFB}}$ under the orthonormal basis $|n\rangle$ that satisfies $a|n\rangle = \sqrt{n}|n-1\rangle$ and $a^\dagger|n\rangle = \sqrt{n+1}|n+1\rangle$. The key to the exact solution is the following substitution for the wavefunctions in Eq. (2):

$$\frac{k_x}{\sqrt{k_x^2 + k_y^2}} \to \frac{1}{2}(|n-1\rangle + |n+1\rangle), \quad \frac{k_y}{\sqrt{k_x^2 + k_y^2}} \to \frac{i}{2}(|n-1\rangle - |n+1\rangle). \quad (3)$$

Under the substituted basis $|n\rangle_0$ and $|n\rangle_1$, $\widetilde{H}_{\text{SFB}}$ is block diagonalized with respect to the quantum number $n$ (see details in SI). The resulting exact expressions of LLs are

$$E_{0,n} = \frac{t}{2l_B^2} \left\{ (2n+1) + 2\xi\sqrt{1-d^2} - \sqrt{\left[(2n+1) + 2\xi\sqrt{1-d^2}\right]^2 + 3d^2} \right\} \text{ for } n \geq 1, \quad (4)$$

$$E_{1,n} = \frac{t}{2l_B^2} \left\{ (2n+1) + 2\xi\sqrt{1-d^2} + \sqrt{\left[(2n+1) + 2\xi\sqrt{1-d^2}\right]^2 + 3d^2} \right\} \text{ for } n \geq -1. \quad (5)$$

$E_{0,n}$ and $E_{1,n}$ comes from the FB and the PB respectively. In Eq. (5), the LLs far away from the BCP ($n \to \infty$) becomes conventional with equal energy spacing $\hbar\omega_c = 2t/l_B^2$, which is defined as the unit of energy. Fig. 2b shows the two branches of LL spectra with $d = 0.2$, highlighting the exact solutions of Eq. (4) for $\xi = 1$ (blue) and $\xi = -1$ (orange) respectively. We note the two black LLs in the right panel come from the PB [Eq. (5)], which is resolved from the plots of a few lowest LLs of Eq. (5) with respect to

$d$ and $\xi$ (see Fig. S1). For $\xi = -1$, the two LLs from the PB drop close to zero energy when $d \to 0$ (dashed lines in Fig. S1). In previous numerical studies, the $\xi = 1$ branch has been discussed extensively[9], while the spectrum of the $\xi = -1$ branch is illustrated in Fig. 6 of ref.[33] without further discussion. Here we show that the two branches originate from the intrinsic chirality of the FB wavefunction [Eq. (2)]. Since the cyclotron motion of an electron under magnetic field also has a chirality $\xi_B = \pm 1$, the production $\xi \cdot \xi_B = \pm 1$ determines which branch to be observed, and a simple way to verify the two branches is by reversing the direction of magnetic field. From a symmetry perspective, the chirality of SFB for $0 < d < 1$ originates from breaking time reversal symmetry in Eq. (1). This has been explicitly illustrated in lattice models by adding phase factors to the hopping parameters[34].

Quantitatively, we define the LL spreading of the SFB as $\Delta = E_{0,\infty} - E_{0,1}$[9], having the following exact expression:

$$\frac{\Delta}{\hbar \omega_c} = -\frac{1}{4}\left[\left(3 + 2\xi\sqrt{1-d^2}\right) - \sqrt{\left(3 + 2\xi\sqrt{1-d^2}\right)^2 + 3d^2}\right]. \quad (6)$$

Fig. 2a shows the two branches of $\Delta$ depending on $d$ for $\xi = \pm 1$. The blue branch with $\xi = 1$ coincides with the numerical result in ref.[9]. Note that the two branches merge at the two endpoints for $d = 0$ and 1. For the non-singular case $d = 0$, the FB wavefunction apparently has no chirality; for the maximally singular case $d = 1$, the Hamiltonians under magnetic field with $\xi = \pm 1$ is related by a unitary transformation.

**LL spectrum of diatomic kagome lattice**
From a material perspective, SFBs with $d = 1$ can be realized in some candidate materials of organic frameworks[35,36], including the superatomic graphene with $sp^2$-like molecular orbitals on honeycomb lattice[18,19], which can be effectively described by a diatomic kagome lattice model in Fig. 1a. It can be viewed by replacing each site of a kagome lattice with a dumbbell of two atoms with tunable bond length $r$ (upper panel of Fig. 3b-e), where we denote the maximal bond length as $r_0$ and define the dimensionless real-space distance $\alpha = r/r_0 \in (0,1)$. We consider a tight-binding (TB) model with the nearest-neighbor (NN), 2$^{nd}$ NN (2NN) and 3NN hoppings labeled in Fig. 1a as $t_1 = 1$, $t_2 = 0$ (bipartite lattice condition)[37] and $t_3 = 0.3$. Compared with kagome lattice, the diatomic kagome lattice has two sets of kagome bands with opposite signs of hoppings[18] and particle-hole symmetry (Fig. 1b). The two SFBs marked red near the $\Gamma$ point can be characterized by Eq. (1) with $d = 1$ and $t = \pm t_3/4$, as obtained by resolvent perturbation theory[9,38] (SI).

Without losing generality, we consider a system with fixed hoppings $t_{1\sim 3}$ and tunable real-space distance $\alpha$. While the band structure keeps exactly the same as Fig. 1b for different $\alpha$, the wavefunctions and hence the quantum geometries of the SFBs will evolve slightly[39,40]. Fig. 1c shows the distribution of quantum metric $g_{\mu\nu}$ of the upper or lower SFB for $\alpha \to 0$. The principal axes of $g_{\mu\nu}$ lie along the tangential and radial directions of a circle drawn around the $\Gamma$ point, which are illustrated by the characteristic ellipses. Closer to $\Gamma$, $g_{\mu\nu}$ is more dominated by the tangential component, which diverges at the singular BCP. For $\alpha \in (0,1)$, $g_{\mu\nu}$ marginally depends on $\alpha$ (see Fig. S2). However, the asymptotic behavior of $g_{\mu\nu}$ close to the BCP is independent of $\alpha$, which is characterized by the maximal quantum distance $d = 1$. The Berry curvature of the FBs, which is proportional to the imaginary part of QGT, vanishes everywhere excluding the singular BCP. Thus, from the quantum geometry perspective, it appears that the SFBs near the $\Gamma$ point in diatomic kagome lattice show little difference from kagome lattice.

Now we calculate the LLs of the diatomic kagome lattice with varying $\alpha$ and fixed hopping amplitudes under commensurate magnetic flux $\phi = 1/50\ \phi_0$, $\phi_0$ is the flux quanta (SI). Generally, the LL spectrum in Fig. 3a is composed of two sets of kagome LLs with opposite sign of energy, respecting the particle-hole symmetry. However, in a zoomed-in view, there is clearly an unexpected evolution of LL spectrum, as shown for the LLs from the upper SFB (Fig. 3b-e). At the limit of $\alpha \to 0$, the ALL spread $\Delta$ has the largest value $\Delta_0$, same as in kagome lattice with NN hopping $t_3$, which can be well described by the 2-band Hamiltonian [Eq. (1)] under a weak magnetic field with $\Delta_0$ given by Eq. (6) for $d = 1$: $\Delta_0/\hbar\omega_c = (2\sqrt{3} - 3)/4 \approx 0.116$. However, when $\alpha$ increases, $\Delta$ is gradually compressed to zero at the limit of $\alpha \to 1$, even though $d$ remains unchanged, significantly beyond the original 2-band effective model[9]. It indicates that the ALL spectrum sensitively depends on the real-space geometry, in addition to the quantum geometry.

**Magnetic disruption of destructive interference**
The evolution of the ALL spreading of the diatomic kagome lattice in Fig. 3 can be intuitively understood by the effect of magnetic field on destructive quantum interference. The wavefunctions of SFBs in this lattice can be expressed as CLSs with alternating phases at the 12 corners of a dodecagon[18] represented by the red and blue dots respectively, as shown in the upper panel of Fig. 4. The dodecagon is composed of six dumbbells with the diatomic distance $r$. For the upper SFB, the phases are the same within each dumbbell, indicating the anti-bonding nature between the two kagome sublattices; the phases change sign for neighboring dumbbells. Apparently, an electron cannot hop out of the CLS due to the destructive interference, of which one leading channel is illustrated by the arrows with hopping amplitude $t_3$. Note that there are other hopping channels which also form perfect destructive interference.

Under magnetic field, the hoppings are modulated by a phase factor $e^{i\theta}$ so that the total phase of a closed loop is proportional to the magnetic flux through its area. For example, the two hoppings marked by arrows in Fig. 4 become $t_3 e^{i\theta_1}$ and $t_3 e^{i\theta_2}$ respectively. Then the destructive interference is disrupted by the magnetic field, leading to broadening of the FBs into the ALL spectrum. From the evolution of ALL spread in Fig. 4a-d, one sees the strength of the disruption is the strongest for $\alpha \to 0$ and gradually reduces to zero for $\alpha \to 1$. This is consistent with the two hopping paths in Fig. 4 becoming closer to each other as $\alpha$ increases, so that the phase difference

$\theta_1 - \theta_2$ becomes smaller (provided that the gauge is smooth). At the $\alpha \to 1$ limit, the two hopping paths coincide and $\theta_1 = \theta_2$. Consequently, the destructive interference is reconstructed to be perfect in the presence of magnetic field, resulting in vanishing $\Delta$ (see SI for the detailed construction of CLS under magnetic field and its destructive interference).

**Semiclassical picture**

The LLs of a Bloch electron under magnetic field can be understood from a semiclassical picture based on the wavepacket dynamics[23]. For one of the SFBs in a diatomic kagome lattice, we may still consider a 2-band effective Hamiltonian describing a wavepacket constructed from both the FB and the PB coupled with magnetic field $B$. Extending Eq. (1), due to the presence of four other bands, one needs to further include the interaction of the wavepacket with the other bands. The leading term in $B$ is the *non-Abelian* orbital magnetic moment[23,41-43]:

$$M_{mn} = -\frac{ie}{2\hbar} \sum_l \left( \varepsilon_l - \frac{\varepsilon_m + \varepsilon_n}{2} \right) (\mathbf{R}_{ml} \times \mathbf{R}_{ln}), \tag{7}$$

where $m$, $n$, $l$ are the band indices. $m, n = 0, 1$ denote the band indices within the SFB nearly-degenerate subspace; for the dominant contribution to $M_{mn}$ near the $\Gamma$ point, $l$ includes the other four bands which are energetically well separated from the SFB subspace. $\varepsilon_l$ is the band energy for $l$ and $\mathbf{R}_{ml} = i\langle u_m | \frac{\partial u_l}{\partial \mathbf{k}} \rangle$ is the interband Berry connection. $M_{mn}$ is a function of $\mathbf{k}$ and gauge-covariant. For the upper SFB, we calculate $M_{mn}$ near the $\Gamma$ point, keeping only the $\mathbf{k}$-independent leading term (SI):

$$M_{mn} = -\frac{et}{2\hbar} \alpha^2 \begin{pmatrix} 0 & 1 \\ 1 & 0 \end{pmatrix}. \tag{8}$$

The diagonal components, which are the Abelian orbital magnetic moment of the FB and PB respectively, must vanish due to the coexistence of inversion and time-reversal symmetry in the absence of spin-orbit coupling[23]. This makes our case different from ref.[44]. The orbital magnetic term $H'_{\text{SFB}} = -B \cdot M_{mn}$ is added to $H_{\text{SFB}}$ in Eq. (1) after being transformed to the same basis. Following the aforementioned block-diagonalization procedure, for $d = 1$ the energy levels are (SI)

$$E_{0,n} = \frac{t}{l_B^2} \left[ n + 1/2 - \sqrt{n(n+1) + (1 - \alpha^2/2)^2} \right] \text{ for } n \geq 1, \tag{9}$$

$$E_{1,n} = \frac{t}{l_B^2} \left[ n + 1/2 + \sqrt{n(n+1) + (1 - \alpha^2/2)^2} \right] \text{ for } n \geq -1. \tag{10}$$

The ALL spectra $E_{0,n}$ (with a constant energy shift $t_1 - 2t_3 = 0.4$) are plotted as the red lines in Fig. 4, which agree very well with the black spectra obtained from the full lattice calculations. Note that the results are obtained under a weaker magnetic field $\phi = 1/100 \, \phi_0$. For strong magnetic field, the lattice effect becomes more prominent, and the higher-order terms must be included. For the upper SFB, from Eq. (9), the dependence of ALL spread $\Delta$ on the dimensionless real-space distance $\alpha$ is

$$\frac{\Delta}{\hbar\omega_c} = -\frac{1}{2}\left[\frac{3}{2} - \sqrt{2 + (1 - \alpha^2/2)^2}\right], \tag{11}$$

which is plotted in Fig. S6. The analysis for the lower SFB is similar, related by the particle-hole symmetry.

The emergence of orbital magnetic moment shares similar physics as the origin of electron spin from the Dirac equation[21-23]. For a Dirac wavepacket constructed from the twofold-degenerate positive-energy branch of the Dirac spectrum, its interaction with the antiparticle branch, which is also described by Eq. (7), gives rise to the self-rotation of the wavepacket to generate the intrinsic magnetic moment of an electron[21]. Here, the "particle and antiparticle" branches are created by the bipartite lattice structure: the bonding and anti-bonding states between the two kagome sublattices, respecting the particle-hole symmetry[18]. The interaction between the two branches generates $M_{mn}$ responding to magnetic field, and most interestingly, it is tunable by the real-space distance $\alpha$ [Eq. (8)]. Similar to Berry curvature being related to the imaginary part of QGT[28,45], the orbital magnetic moment corresponds to the imaginary part of so-called quasi-QGT[45,46]. Although the real part of QGT ($g_{\mu\nu}$ and $d$) is independent of $\alpha$ near the singular BCP (SI), the imaginary part of quasi-QGT (or its non-Abelian generalization) encodes phase information and sensitively depends on $\alpha$. Consequently, a unique scenario occurs that the magnetic response is governed by both $d$ and $\alpha$, showing an intriguing interplay of quantum and real-space geometry.

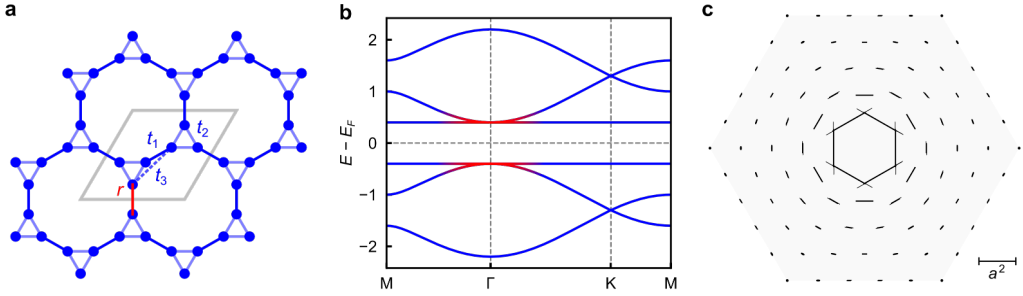

**Fig. 1 | The real-space geometry, band structure and quantum geometry of a diatomic kagome lattice. a**, The diatomic kagome lattice with two atoms like a dumbbell on each lattice site. The bond length of the dumbbell (marked red) is $r$. The intra-dumbbell, inter-dumbbell and cross-dumbbell hopping are labeled as $t_1$, $t_2$ and $t_3$, respectively. For $t_2 = 0$, it is a bipartite lattice with chiral symmetry. The grey rhombus marks the unit cell. **b**, The band structure with yin-yang FBs around the Fermi level obtained by setting $t_1 = 1$, $t_2 = 0$ and $t_3 = 0.3$, exhibiting particle-hole symmetry. The bands marked red highlight the singular BCPs at the $\Gamma$ point between the SFB and a parabolic band, accounting for the ALLs. Note that the band structure is independent of $r$ explicitly. **c**, The quantum metric distribution for $r \to 0$. The ellipses represent the principal axes and components of quantum metric tensor. The grey shadow marks the first Brillouin zone (BZ). $a$ is the lattice constant. Note that the quantum metric marginally depends on $r$ (see SI).

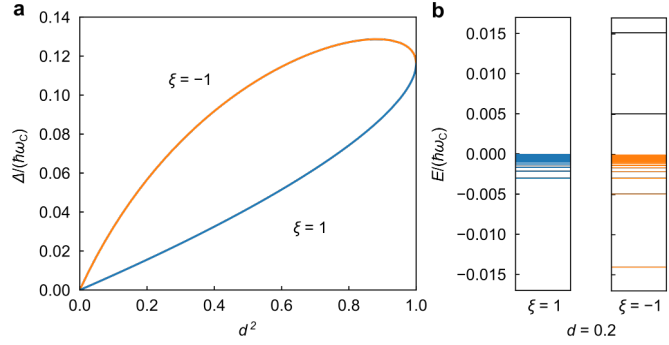

**Fig. 2 | The exact solutions of two-branch Landau level spectra. a**, The plot of the dimensionless LL spreading $\Delta/\hbar\omega_c$ as a function of the maximal quantum distance $d$. **b**, The two different LL spectrum for $d = 0.2$. The blue and orange color corresponds to the $\xi = \pm 1$ branch respectively. The two black LLs in the right panel come from the parabolic band.

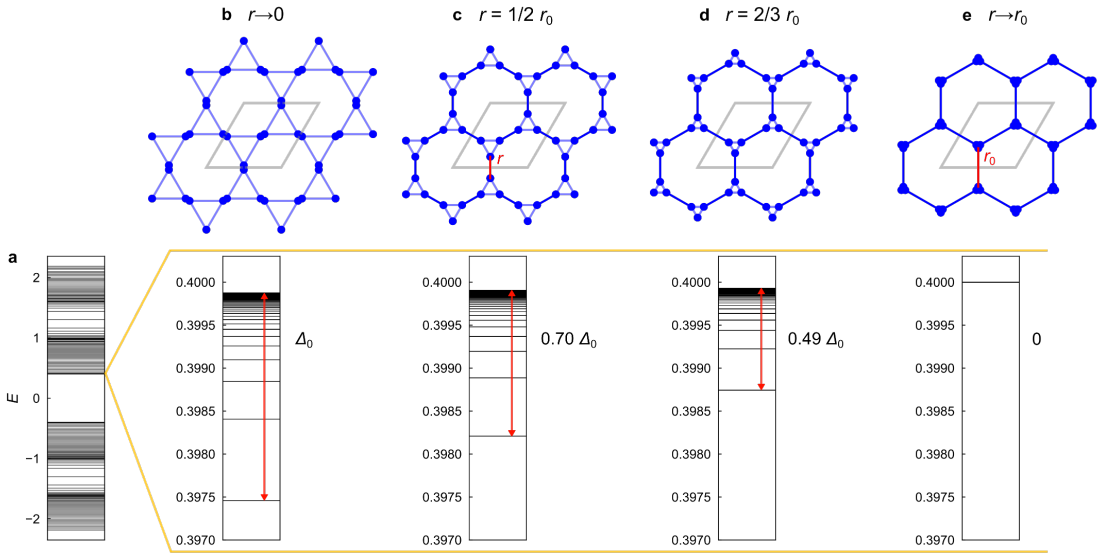

**Fig. 3 | The evolution of ALLs of diatomic kagome lattice. a**, The LLs calculated using the magnetic flux $\phi = 1/50\,\phi_0$ for $r \to 0$, $\phi_0$ is the flux quanta. **b-e**, The evolution of ALLs of the conduction SFB (see Fig. 1b) with varying $r \in (0, r_0)$ in a zoomed-in view. The red arrows mark the ALL spreading $\Delta$. In e $\Delta$ vanishes. $\Delta_0$ is the same as the ALL spreading of the SFB in a kagome lattice with the NN hopping $t_3$ under the same flux[9].

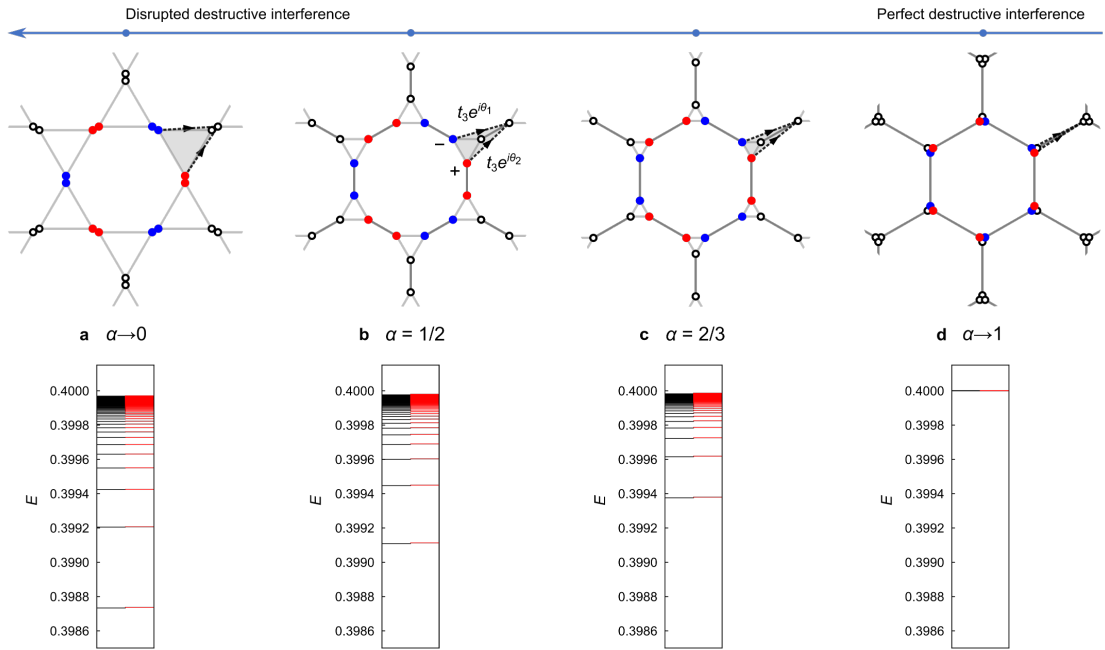

**Fig. 4 | The disrupted destructive interference and comparison between the ALLs obtained from the TB lattice model and two-band effective model. a-d**, The upper panel shows the gradual disruption of the destructive interference induced by magnetic field. The red and blue dots represent the positive and negative phase of the SFB wavefunction at the corners of the dodecagonal CLS respectively. The arrows show two representative $t_3$ hoppings out of the CLS which are modulated by a phase factor $e^{i\theta_1}$ or $e^{i\theta_2}$ under magnetic field, respectively, as indicated in **b**. **d**, The destructive interference is reconstructed to be perfect in the presence of magnetic field. The lower panel compares the ALLs calculated from the 6-band TB model (black lines) and two-band effective model (red lines). The calculation setup is the same as Fig. 3 but with $\phi = 1/100\, \phi_0$. $\alpha$ is defined as $r/r_0$ (see Fig.3).

## Data availability

The data that support the findings of this study are included in this article and its supplementary information file and are available from the corresponding author upon reasonable request.

## Code availability

The codes used to generate the data of the current study are available from the corresponding author on reasonable request.

## Acknowledgements

We thank Guangjie Li, Zheng Liu, Feng Qi, Yongshi Wu, and Mengxing Ye for helpful discussions. This work was fully supported by the DOE-BES (No. DE-FG02-04ER46148).

## Author contributions

F.L. conceived the project; X.L. did the calculations; X.L. and F.L. analyzed the calculation results; X.L. and F.L. wrote the manuscript.

## Competing interests

The authors declare no competing interests.

# Supplementary information for "Interplay of quantum and real-space geometry in the anomalous Landau levels of singular flat bands"

Xuanyu Long[1] and Feng Liu[1*]

[1]Department of Materials Science and Engineering, University of Utah, Salt Lake City, Utah 84112,

USA

*To whom correspondence should be addressed: fliu@ eng.utah.edu

## I. Block diagonalization of the 2-band Hamiltonian

In this section, we show how to diagonalize the 2-band effective Hamiltonian [Eq. (1) in the main text]

$$H_{\text{SFB}} = t \begin{pmatrix} k_x^2 + (1-d^2)k_y^2 & d\sqrt{1-d^2}k_y^2 - i\xi d k_x k_y \\ d\sqrt{1-d^2}k_y^2 + i\xi d k_x k_y & d^2 k_y^2 \end{pmatrix} \quad \text{(S1)}$$

under magnetic field in detail, where $d = d_{\max}$ (the subscript is emitted for simplicity) is the maximal quantum distance near the band touching point, $\xi = \pm 1$ is the chirality of the singular flat band (SFB) wavefunction, and $t$ sets the energy scale. Following the procedures in ref.[1-3], we first substitute the lattice momenta $k_x$ and $k_y$ in Eq. (S1) by $k_x \to (a + a^\dagger)/(\sqrt{2}l_B)$ and $k_y \to i(a - a^\dagger)/(\sqrt{2}l_B)$, where $a$ ($a^\dagger$) is the annihilation (creation) operator satisfying $[a, a^\dagger] = 1$ and $l_B = \sqrt{\hbar/eB}$ is the magnetic length. Note that since $k_x$ and $k_y$ are non-commutative, upon the substitution the term $k_x k_y$ is symmetrized as $(k_x k_y + k_y k_x)/2 = i(a^2 - a^{\dagger 2})/(2l_B^2)$. The substituted Hamiltonian is

$$\widetilde{H}_{\text{SFB}} = \frac{t}{2l_B^2} \begin{pmatrix} h_{11} & h_{12} \\ h_{12}^\dagger & h_{22} \end{pmatrix}, \quad \text{(S2)}$$

where

$$\begin{aligned}
h_{11} &= d^2(a^2 + a^{\dagger 2}) + (2 - d^2)(2a^\dagger a + 1), \\
h_{12} &= d\left(\xi - \sqrt{1-d^2}\right) a^2 - d\left(\xi + \sqrt{1-d^2}\right) a^{\dagger 2} + d\sqrt{1-d^2}(2a^\dagger a + 1), \\
h_{22} &= -d^2(a^2 + a^{\dagger 2}) + d^2(2a^\dagger a + 1).
\end{aligned} \quad \text{(S3)}$$

Then we perform the following substitution [Eq. (3) in the main text]:

$$\frac{k_x}{\sqrt{k_x^2 + k_y^2}} \to \frac{1}{2}(|n-1\rangle + |n+1\rangle), \quad \frac{k_y}{\sqrt{k_x^2 + k_y^2}} \to \frac{i}{2}(|n-1\rangle - |n+1\rangle) \quad \text{(S4)}$$

for the wavefunctions [Eq. (2) in the main text]

$$V_0 = \frac{1}{\sqrt{k_x^2 + k_y^2}} \begin{pmatrix} -dk_y \\ i\xi k_x + \sqrt{1-d^2}k_y \end{pmatrix} \text{ and } V_1 = \frac{1}{\sqrt{k_x^2 + k_y^2}} \begin{pmatrix} \sqrt{1-d^2}k_y - i\xi k_x \\ dk_y \end{pmatrix}, \quad (S5)$$

where $|n\rangle$ is the orthonormal basis that satisfies $a|n\rangle = \sqrt{n}|n-1\rangle$ and $a^\dagger|n\rangle = \sqrt{n+1}|n+1\rangle$. The above step is the key to obtain the exact solution. The substituted wavefunctions are

$$
\begin{aligned}
|n\rangle_0 &= \frac{i}{2}\begin{pmatrix} -d|n-1\rangle + d|n+1\rangle \\ \left(\xi + \sqrt{1-d^2}\right)|n-1\rangle + \left(\xi - \sqrt{1-d^2}\right)|n+1\rangle \end{pmatrix}, \\
|n\rangle_1 &= \frac{i}{2}\begin{pmatrix} \left(\sqrt{1-d^2} - \xi\right)|n-1\rangle - \left(\sqrt{1-d^2} + \xi\right)|n+1\rangle \\ d|n-1\rangle - d|n+1\rangle \end{pmatrix}.
\end{aligned}
\quad (S6)
$$

Applying the substituted Hamiltonian $\widetilde{H}_{\mathrm{SFB}}$ onto bases $|n\rangle_0$ and $|n\rangle_1$, after some algebra, we obtain

$$\widetilde{H}_{\mathrm{SFB}}|n\rangle_0 = \frac{i\xi t}{4l_B^2}\begin{pmatrix} d\left(\xi - \sqrt{1-d^2}\right)\left[-(2n-1) + 2\sqrt{n(n+1)}\right]|n-1\rangle + \\ d\left(\xi + \sqrt{1-d^2}\right)\left[(2n+3) - 2\sqrt{n(n+1)}\right]|n+1\rangle \\ d^2\left[(2n-1) - 2\sqrt{n(n+1)}\right]|n-1\rangle + \\ d^2\left[(2n+3) - 2\sqrt{n(n+1)}\right]|n+1\rangle \end{pmatrix},$$

$$\widetilde{H}_{\mathrm{SFB}}|n\rangle_1 = -\frac{i\xi t}{4l_B^2}\begin{pmatrix} \left\{\left[2\left(1 - \xi\sqrt{1-d^2}\right) - d^2\right](2n-1) + 2d^2\sqrt{n(n+1)}\right\}|n-1\rangle + \\ \left\{\left[2\left(1 + \xi\sqrt{1-d^2}\right) - d^2\right](2n+3) + 2d^2\sqrt{n(n+1)}\right\}|n+1\rangle \\ -d\left[\left(\xi - \sqrt{1-d^2}\right)(2n-1) + 2\left(\xi + \sqrt{1-d^2}\right)\sqrt{n(n+1)}\right]|n-1\rangle + \\ d\left[\left(\xi + \sqrt{1-d^2}\right)(2n+3) + 2\left(\xi - \sqrt{1-d^2}\right)\sqrt{n(n+1)}\right]|n+1\rangle \end{pmatrix}.$$

(S7)

It is straightforward to find when $m \neq n$, the matrix elements ${}_i\langle m|\widetilde{H}_{\mathrm{SFB}}|n\rangle_j$ vanish, where $i,j = 0$ or $1$ are the band indices. Thus $\widetilde{H}_{\mathrm{SFB}}$ is block diagonal with respect to the quantum number $n$ under the basis of Eq. (S1). The matrix elements for the $n^{\text{th}}$ $2\times 2$ block $\left[\widetilde{H}_{\mathrm{SFB}}(n)\right]_{ij} = {}_i\langle n|\widetilde{H}_{\mathrm{SFB}}|n\rangle_j$ are given as

$$
\begin{aligned}
{}_0\langle n|\widetilde{H}_{\mathrm{SFB}}|n\rangle_0 &= \frac{t}{l_B^2} \cdot \frac{d^2}{2}\left[(2n+1) - 2\sqrt{n(n+1)}\right], \\
{}_0\langle n|\widetilde{H}_{\mathrm{SFB}}|n\rangle_1 &= -\frac{t}{l_B^2}\left\{\xi d + \frac{d\sqrt{1-d^2}}{2}\left[(2n+1) - 2\sqrt{n(n+1)}\right]\right\}, \\
{}_1\langle n|\widetilde{H}_{\mathrm{SFB}}|n\rangle_1 &= \frac{t}{l_B^2}\left\{(2n+1) + 2\xi\sqrt{1-d^2} - \frac{d^2}{2}\left[(2n+1) - 2\sqrt{n(n+1)}\right]\right\},
\end{aligned}
\quad (S8)
$$

and ${}_1\langle n|\widetilde{H}_{\mathrm{SFB}}|n\rangle_0 = {}_0\langle n|\widetilde{H}_{\mathrm{SFB}}|n\rangle_1$. After diagonalizing the $2\times 2$ matrix $\left[\widetilde{H}_{\mathrm{SFB}}(n)\right]_{ij}$, the

Landau level (LL) spectra are obtained as Eq. (4) and (5) and plotted in Fig. 2b in the main text.

We note that in Fig. 2b, for $\xi = -1$ there are two lowest-energy LLs (distinguished by black color for clarity) from the parabolic band (PB) close to zero energy. The evolution of several lowest LLs from the PB with respect to $d$ for $\xi = \pm 1$ are plotted in Fig. S1 based on Eq. (5) in the main text. One can see for $\xi = -1$ (dashed lines), the two lowest LLs drop close to zero energy when $d \to 0$.

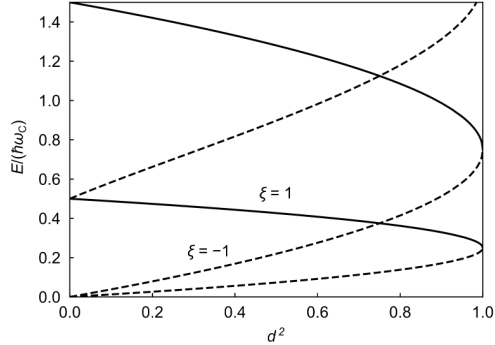

**Fig. S1 | The evolution of several lowest Landau levels from the parabolic band.** The solid and dashed lines correspond to $\xi = \pm 1$ respectively.

## II. The dependence of quantum metric on real-space distance

Here we calculate the quantum geometry of the SFBs in diatomic kagome lattice for varying real-space distance $r \in (0, r_0)$, and show it marginally depends on $r$. In practice, we use the following equation to calculate the quantum geometry tensor numerically[4]:

$$Q_{\mu\nu} = \sum_{n \neq 0} \frac{\langle u_0 | \partial_\mu H | u_n \rangle \langle u_n | \partial_\nu H | u_0 \rangle}{(\varepsilon_0 - \varepsilon_n)^2}, \tag{S9}$$

where $\mu, \nu = x, y$ are the indices of momentum, $\varepsilon_n$ and $|u_n\rangle$ are the energy and wavefunction of band $n$ respectively, and $0$ stands for the upper flat band. The quantum metric $g_{\mu\nu} = \operatorname{Re} Q_{\mu\nu}$ is the real part of Eq. S4. Fig. S2e-h shows the evolution of $g_{\mu\nu}(\mathbf{k})$ illustrated by the characteristic ellipses as a function of $r \in (0, r_0)$. The results for the lower flat band are the same. One sees there is little difference between Fig. S2e-h, and especially, the asymptotic behavior close to the $\Gamma$ point remains the same, which is characterized with $d = 1$. Thus, as a representative, only Fig. S2e is shown in the main text.

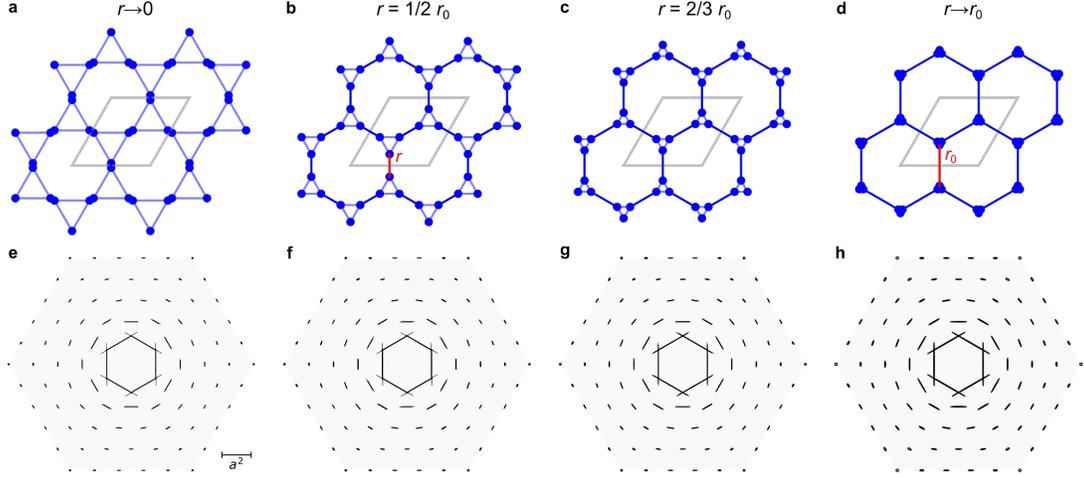

**Fig. S2 | The evolution of real-space and quantum geometry of diatomic kagome lattices. a-d**, The evolution of real-space geometry by varying $r \in (0, r_0)$ (marked red in **b**). The maximum $r_0 = \sqrt{3}/6\, a$ (marked red in **d**), where $a$ is the lattice constant. **e-h**, The evolution of quantum geometry of SFB in correspondence with **a-d** calculated using the bands in Fig. 2b. The grey shadow marks the first Brillouin zone (BZ). The ellipses represent the principal axes and components of the quantum metric tensor.

### III. Calculating Landau levels of the diatomic kagome lattice

Here we show how to construct the magnetic unit cell and calculate the LLs for the diatomic kagome lattice. We consider the simplest case of magnetic flux $\phi = 1/N\, \phi_0$, where $N$ is an integer. The magnetic unit cell is constructed as being elongated along the $x$ direction by $qN$ times, where $q$ is another integer, as sketched in Fig. S3 for the case of $qN = 5$. The $t_1$ and $t_3$ bonds are plotted as the dark and light grey lines respectively, including those crossing the boundary. Under a Landau-type gauge $\mathbf{A} = B(x - y/\sqrt{3})\mathbf{e}_y$, each bond gains a phase $t_{ij} \to t_{ij} e^{i\theta_{ij}}$ determined by the positions of its two end points: $\theta_{ij} = -2\pi\phi/\phi_0 \cdot \Delta y_{ij} \cdot \bar{x}_{ij}/A_0$, where $\Delta y_{ij} = y_j - y_i$, $\bar{x}_{ij} = [(x_i + x_j) - (y_i + y_j)/\sqrt{3}]/2$, $A_0 = \sqrt{3}a^2/2$ and $a$ is the lattice constant. This ensures the total phase of any closed loop (illustrated by the red shadow for example) is proportional to the magnetic flux through its area. The form of $\bar{x}_{ij}$ guarantees the phase of each bond is invariant upon a translation of $\mathbf{e}_2$, i.e., to maintain the translational symmetry along the $\mathbf{e}_2$ direction. For the translational symmetry along the $qN\mathbf{e}_1$ direction, it requires that the total phase of the blue shadowed area across the boundary is the same as the red one. This leads to $q = 2, 4, 6$ and $1$ for $r \to 0$, $r = 1/2\, r_0$, $r = 2/3\, r_0$ and $r \to r_0$, respectively. In practice, we construct a $6qN \times 6qN$ Hamiltonian and only consider the zero-momentum case for simplicity. Since each LLs are $q$-fold quasi-degenerate, we only pick one from each degenerate subspace to plot Fig. 3 and Fig. 4 in the main text.

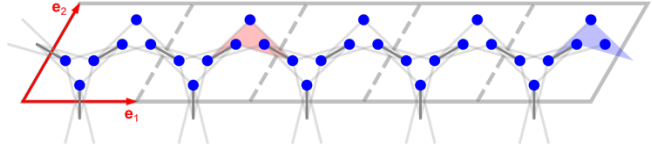

**Fig. S3 | The magnetic unit cell and boundary condition.** The magnetic unit cell is composed of five original unit cells. $\mathbf{e}_1$ and $\mathbf{e}_2$ are the basis vectors of the original unit cell. The dark and light grey lines show all the $t_1$ and $t_3$ bonds respectively. The red shadow marks a closed loop enclosed by three $t_3$ and one $t_1$ bonds. The blue shadow has the same shape as the red one but crosses the boundary.

## IV. The compact localized states under magnetic field

Here we show the destructive interference is reconstructed to be perfect for $r \to r_0$ even under magnetic field. We consider the simplest case of magnetic flux $\phi = 1/N\,\phi_0$, where $N$ is an integer. One compact localized state (CLS) under magnetic field for $N = 4$ is constructed in Fig. S4, of which the size is $N$ times of the unit cell. Under the Landau gauge $\mathbf{A} = Bx\mathbf{e}_y$, the phase of each bond $\theta_{ij}$ is calculated similarly as discussed in Sec. III above. The (unnormalized) wavefunction of the CLS $w_i = \pm e^{i\varphi_i}$ are illustrated by the colormap distributing along a closed loop, where $i$ labels the positions, satisfying $\varphi_j = \varphi_i + \theta_{ij}$ if $i$ and $j$ are linked by a $t_1$ bond marked by grey arrows. For two coinciding sites, their wavefunctions are opposite. We can arrange $\varphi_i$ compatibly since the magnetic flux through the closed loop formed by all $t_1$ bonds equals $\phi_0$, the flux quanta.

One can find this CLS is an eigenstate of the Hamiltonian under magnetic field. The dashed black arrows mark two pairs of destructive interference. We have shown that the right pair forms destructive interference in the main text. For the left one, the wavefunctions at its two starting points are $e^{i\varphi_1}$ and $-e^{i\varphi_3}$ respectively, where $\varphi_3 = \varphi_1 + \theta_{12} + \theta_{23}$. For $r \to r_0$, the left and right $t_3$ hoppings coincide with the $t_1$ hoppings, having a phase $e^{i\theta_{12}}$ and $e^{-i\theta_{23}}$ respectively. Since $e^{i\varphi_1}e^{i\theta_{12}} - e^{i\varphi_3}e^{-i\theta_{23}} = 0$, the two paths form perfect destructive interference. Note that all of the above analyses hold for any gauge and any shape of CLS that has a size of $N$ times of the unit cell.

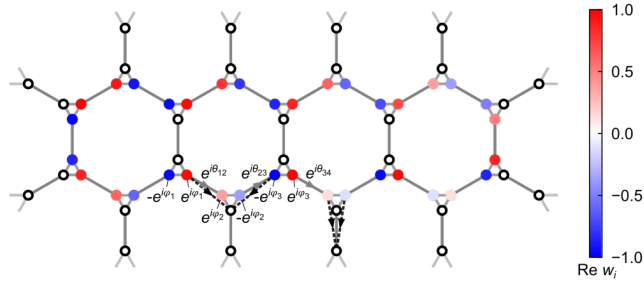

**Fig. S4 | The compact localized state and its destructive interference under magnetic field.** The colormap shows the wavefunction distribution $w_i = \pm e^{i\varphi_i}$ obtained under the Landau gauge, where $i$

labels the positions. The grey arrows show three $t_1$ bonds. The dashed black arrows show two pairs of destructive interference. Note that $r$ is plotted slightly away from $r_0$ (see Fig. S2a-d) to show the wavefunction and its destructive interference clearly.

## V. Calculating the effective Hamiltonian and the Non-Abelian orbital magnetic moment

In this section, we show how to calculate the 2-band effective Hamiltonian and the non-Abelian orbital magnetic moment from the 6-band tight-binding model in a unified framework based on the resolvent perturbation theory. Since both the effective Hamiltonian and the non-Abelian orbital magnetic moment [Eq. (1) and (8) in the main text] are gauge covariant, here we show explicitly under which basis they are calculated.

We first write down the 6-band tight-binding Hamiltonian $H$ of the diatomic kagome lattice depicted in Fig. S1. We only consider the $t_1$ and $t_3$ bonds (see Fig. 1a in the main text, the exact chiral limit). We have $H_{14} = t_1 e^{2id_2}$, $H_{15} = t_3 e^{i(d_2+a_1/2+d_1)} + t_3 e^{i(d_2-a_1/2+d_1)}$, $H_{16} = t_3 e^{i(d_2+a_2/2+d_3)} + t_3 e^{i(d_2-a_2/2+d_3)}$, $H_{24} = t_3 e^{i(d_1+a_1/2+d_2)} + t_3 e^{i(d_1-a_1/2+d_2)}$, $H_{25} = t_1 e^{2id_1}$, $H_{26} = t_3 e^{i(d_1+a_3/2+d_3)} + t_3 e^{i(d_1-a_3/2+d_3)}$, $H_{34} = t_3 e^{i(d_3+a_2/2+d_2)} + t_3 e^{i(d_3-a_2/2+d_2)}$, $H_{35} = t_3 e^{i(d_3+a_3/2+d_1)} + t_3 e^{i(d_3-a_3/2+d_1)}$, $H_{36} = t_1 e^{2id_3}$ and $H_{ij} = H_{ji}^*$, where $d_i = \mathbf{k} \cdot \mathbf{d}_i$ and $a_i = \mathbf{k} \cdot \mathbf{a}_i$ for $i = 1-3$.

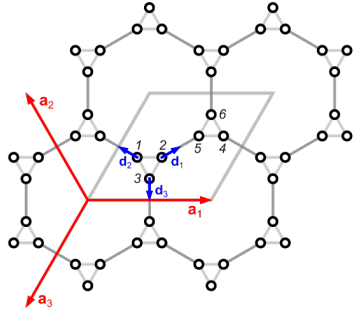

**Fig. S5 | The structure of diatomic kagome lattice and some conventions.** $\mathbf{a}_{1-3}$ and $\mathbf{d}_{1-3}$ are the structural vectors. The numbers 1-6 label the atomic sites.

At the $\Gamma$ point, $H$ can be diagonalized analytically using the unitary transformation $\widetilde{H} = U_0^\dagger H U_0$, where

$$U_0 = \begin{pmatrix} \frac{1}{\sqrt{6}} & \frac{1}{2} & -\frac{1}{\sqrt{12}} & \frac{1}{2} & -\frac{1}{\sqrt{12}} & \frac{1}{\sqrt{6}} \\ \frac{1}{\sqrt{6}} & -\frac{1}{2} & -\frac{1}{\sqrt{12}} & -\frac{1}{2} & -\frac{1}{\sqrt{12}} & \frac{1}{\sqrt{6}} \\ \frac{1}{\sqrt{6}} & 0 & \frac{2}{\sqrt{12}} & 0 & \frac{2}{\sqrt{12}} & \frac{1}{\sqrt{6}} \\ -\frac{1}{\sqrt{6}} & -\frac{1}{2} & \frac{1}{\sqrt{12}} & \frac{1}{2} & -\frac{1}{\sqrt{12}} & \frac{1}{\sqrt{6}} \\ -\frac{1}{\sqrt{6}} & \frac{1}{2} & \frac{1}{\sqrt{12}} & -\frac{1}{2} & -\frac{1}{\sqrt{12}} & \frac{1}{\sqrt{6}} \\ -\frac{1}{\sqrt{6}} & 0 & -\frac{2}{\sqrt{12}} & 0 & \frac{2}{\sqrt{12}} & \frac{1}{\sqrt{6}} \end{pmatrix}. \tag{S10}$$

The energy levels are $-(t_1 + 4t_3)$, $-(t_1 + 2t_3)$, $-(t_1 + 2t_3)$, $t_1 + 2t_3$, $t_1 + 2t_3$, and $t_1 + 4t_3$ respectively (see band structure in Fig. 1b in the main text). The columns of $U_0$ corresponds to the wavefunctions $v_{1-6}$ at the $\Gamma$ point. Note that for $v_{1-3}$, the wavefunction on sites 4-6 has an opposite sign of that of sites 1-3, corresponding to the low-energy bonding states (valence bands). While for $v_{4-6}$, the wavefunction on sites 4-6 has the same sign of that of sites 1-3, corresponding to the high-energy antibonding states (conduction bands). Then we construct a 2-band effective Hamiltonian for the upper SFB, which can be obtained by the resolvent perturbation theory[2,5]:

$$H_{\text{eff}} = PHP + PHQ \frac{1}{\varepsilon - QHQ} QHP. \tag{S11}$$

$P$ is the projection operator onto the subspace spanned by $v_4$ and $v_5$, $Q = 1 - P$ and $\varepsilon = t_1 + 2t_3$. In practice, we first perform a unitary transformation $\tilde{H} = U_0^\dagger H U_0$ and let $P = \text{Diag}(0,0,0,1,1,0)$, then apply Eq. (S11) to $\tilde{H}$. After expanding to the quadratic order of $\boldsymbol{k}$ and dropping the constant term $t_1 + 2t_3$, an effective Hamiltonian is obtained as

$$H_{\text{eff}} = t \begin{pmatrix} k_x^2 & k_x k_y \\ k_x k_y & k_y^2 \end{pmatrix}. \tag{S12}$$

Here $t = t_3/4$. Note that the $H_{\text{eff}}$ matrix explicitly depends on the gauge choice of its bases, which are defined as $v_4$ and $v_5$ for now. If one applies a $2 \times 2$ unitary transformation to $v_4$ and $v_5$, the results will change accordingly.

Now we calculate the non-Abelian orbital magnetic moment for the upper SFB starting from the following expression [Eq. (7) in the main text]:

$$M_{mn} = -\frac{ie}{2\hbar} \sum_l \left( \varepsilon_l - \frac{\varepsilon_m + \varepsilon_n}{2} \right) (\boldsymbol{R}_{ml} \times \boldsymbol{R}_{ln}). \tag{S13}$$

The interband Berry connection is defined as $\boldsymbol{R}_{ml} = i\langle u_m | \frac{\partial u_l}{\partial \boldsymbol{k}} \rangle$ for $m \neq l$. To avoid the

explicit derivative of the wavefunctions, we reformulate $\boldsymbol{R}_{ml}$ based on the identity $\left\langle u_m \left| \frac{\partial H}{\partial \boldsymbol{k}} \right| u_l \right\rangle = (\varepsilon_l - \varepsilon_m) \langle u_m | \frac{\partial u_l}{\partial \boldsymbol{k}} \rangle$:

$$\boldsymbol{R}_{ml} = \frac{i \left\langle u_m \left| \frac{\partial H}{\partial \boldsymbol{k}} \right| u_l \right\rangle}{\varepsilon_l - \varepsilon_m}. \tag{S14}$$

Since we only focus on the $\boldsymbol{k}$-independent leading term near the $\Gamma$ point, a further approximation is made that all the terms in Eq. (S13) and Eq. (S14) are evaluated at the $\Gamma$ point:

$$M_{mn} = \frac{ie}{2\hbar} \sum_l \frac{1}{\varepsilon_m^0 - \varepsilon_l^0} \left( \left\langle u_m^0 \left| \left(\frac{\partial H}{\partial \boldsymbol{k}}\right)_0 \right| u_l^0 \right\rangle \times \left\langle u_l^0 \left| \left(\frac{\partial H}{\partial \boldsymbol{k}}\right)_0 \right| u_n^0 \right\rangle \right), \tag{S15}$$

where the sub- and super-script "0" means evaluated at the $\Gamma$ point and $\varepsilon_m^0 = \varepsilon_n^0 = \varepsilon$. $|u_m^0\rangle$ and $|u_l^0\rangle$ are the columns of $U_0$. Eq. (S15) can be written in a more compact form

$$M_{mn} = \frac{ie}{2\hbar} [P \left(\frac{\partial \widetilde{H}}{\partial k_x}\right)_0 Q \frac{1}{\varepsilon - Q\widetilde{H}_0 Q} Q \left(\frac{\partial \widetilde{H}}{\partial k_y}\right)_0 P - (x \leftrightarrow y)]. \tag{S16}$$

This equation can be intuitively understood from the non-commutativity of $k_x$ and $k_y$ under magnetic field: $[k_x, k_y] = -i\frac{e}{\hbar}B$. The $k_x k_y$ term in the effective Hamiltonian Eq. (S12) comes from Eq. (S11) applied to $\widetilde{H}$. Now we trace back its origin in detail, focusing on the difference between $k_x k_y$ and $k_y k_x$. Some of the terms directly come from $\widetilde{H}$, which has a symmetric contribution of $k_x k_y$ and $k_y k_x$. However, for the production of three terms $P\widetilde{H}Q$, $\frac{1}{\varepsilon - Q\widetilde{H}Q}$ and $Q\widetilde{H}P$, if one term contributes $k_x$, another contributes $k_y$, and the third contributes a non-zero constant, there will be asymmetric contributions of $k_x k_y$ and $k_y k_x$. Since $P\widetilde{H}_0 Q = Q\widetilde{H}_0 P = 0$, the constant term must come from $\frac{1}{\varepsilon - Q\widetilde{H}Q}$. Then either $P\widetilde{H}Q$ contributes $k_x$ and $Q\widetilde{H}P$ contributes $k_y$, or *vice versa*. The anti-symmetric contribution to the effective Hamiltonian, which is proportional to $(k_x k_y - k_y k_x)/2 = -\frac{ie}{2\hbar}B$, can be written as $H'_{\text{eff}} = -B \cdot M$ and $M$ is given by Eq. (S16). The above analysis highlights the spirit of "commutator expansion" proposed by Kohn[6] which can be treated rigorously by using Moyal product[7-10]. Eq. (S11) and Eq. (S16) enable a systematic extraction of the effective Hamiltonian and the orbital magnetic moment in a unified framework, representing a "nontrivial generalization of Löwdin partitioning techniques" discussed in ref.[11].

From Eq. (S16), the non-Abelian orbital magnetic moment is obtained:

$$M_{mn} = \frac{et}{2\hbar} \alpha^2 \begin{pmatrix} 0 & i \\ -i & 0 \end{pmatrix}, \tag{S17}$$

which is written under the same basis as Eq. (S12) and $\alpha = r/r_0$. Note that when $d = 1$ and $\xi = 1$, the effective Hamiltonian Eq. (S1) [Eq. (1) in the main text] becomes

$$H_{\text{SFB}} = t \begin{pmatrix} k_x^2 & -ik_x k_y \\ ik_x k_y & k_y^2 \end{pmatrix}, \tag{S18}$$

which can be related to Eq. (S12) by a unitary transformation $H_{\text{SFB}} = U_1^\dagger H_{\text{eff}} U_1$, where $U_1 = \text{Diag}(i, 1)$. Then $M_{mn}$ accordingly becomes

$$M_{mn} = \frac{et}{2\hbar} \alpha^2 \begin{pmatrix} 0 & 1 \\ 1 & 0 \end{pmatrix}. \tag{S19}$$

Finally, we transform $M_{mn}$ to the band basis. Eq. (S18) can be diagonalized by a unitary transformation $U_2^\dagger H_{\text{SFB}} U_2$, and

$$U_2 = \frac{1}{\sqrt{k_x^2 + k_y^2}} \begin{pmatrix} -k_y & -ik_x \\ ik_x & k_y \end{pmatrix}. \tag{S20}$$

The columns of $U_2$ are given by the wavefunctions Eq. (S5). Applying $U_2$ to Eq. (S19), we obtain Eq. (8) in the main text.

## VI. The dependence of anomalous Landau level spreading on real-space distance

For $d = 1$, we can also block diagonalize the 2-band effective Hamiltonian analytically even taking into account the non-Abelian orbital magnetic moment. We set $\xi = 1$ and then $H_{\text{SFB}}$ is given by Eq. (S12). There is an additional orbital magnetic moment contribution $H'_{\text{SFB}} = -B \cdot M_{mn}$ determined by Eq. (S13). Following the steps in Sec. I, we substitute number "1" in $H'_{\text{SFB}}$ by identity matrix in the $|n\rangle$ space and obtain $\widetilde{H}'_{\text{SFB}}$. We find that $\widetilde{H}'_{\text{SFB}}$ is also block diagonalized with respect to the quantum number $n$ under the basis of Eq. (S1). The matrix elements for the $n^{\text{th}}$ 2×2 block $[\widetilde{H}'_{\text{SFB}}(n)]_{ij} = {}_i\langle n|\widetilde{H}'_{\text{SFB}}|n\rangle_j$ are ${}_1\langle n|\widetilde{H}'_{\text{SFB}}|n\rangle_0 = {}_0\langle n|\widetilde{H}'_{\text{SFB}}|n\rangle_1 = \frac{t}{l_B^2} \cdot \frac{\alpha^2}{2}$ and ${}_0\langle n|\widetilde{H}'_{\text{SFB}}|n\rangle_0 = {}_1\langle n|\widetilde{H}'_{\text{SFB}}|n\rangle_1 = 0$. Combining with Eq. (S3), the LL spectra Eq. (9) and (10) in the main text are obtained by diagonalizing each $2 \times 2$ block. The dependence of ALL spreading $\Delta$ on dimensionless real-space distance $\alpha$ is obtained as Eq. (11) in the main text, which is plotted in Fig. S6.

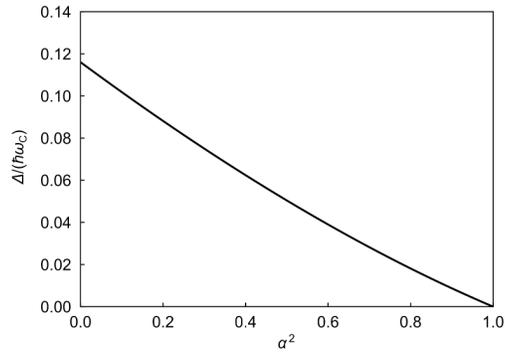

**Fig. S6 | The dependence of anomalous Landau level spreading on real-space geometry.** $\alpha$ is the real-space dimensionless distance defined as $r/r_0$.